# THERMODYNAMIC DESCRIPTION OF SYSTEMS Cd-Te , Hg-Te AND Cd-Hg-Te USING THE MODEL OF ASSOCIETED LIQUID SOLUTION

**A. Halimi , M. S. Ferah**

*Laboratory of physics and chemistry of semiconductors (LPCS),*

*Department of physics,University of Mentouri-Constantine, 25000 Algeria.*

*Corresponding authors. E-mail : h_ali_mi@yahoo.fr tel/fax :(213031)614711*

**Abstract**. This study is devoted in the search of a model describing the best possible liquid and solid phases of the systems Cd-Te, Hg-Te, and Cd-Hg-Te. For the liquid phases, we used the model of Sub-regular Associated Solution (S.A.S). Thermodynamic quantities of dissociation, for the associated species are calculated by the numerical resolution of the phase equilibrium equations. The results of this calculation indicate a more marked associated character in Cd-Te than in Hg-Te with a pseudo-regular behavior of the binary liquid Cd-Te. The phases diagrams calculated from thermodynamic quantities, describe very well the experimental diagrams especially for the binary systems. This models used to calculate the parameters of interactions, enthalpy and entropy of dissociation and mixture, this thermodynamic quantities not measurable used to describe phases diagrams and other studies.

## 1. INTRODUCTION

the experimental data as the quantity of formation and fusion, the partial pressures of the vapors of saturating gas, and the experimental phases diagrams, experimental methods are more difficult, costly and ever complete. the description of the systems to several phases requires the use of theoretical thermodynamic models generally for the liquid phase and the solid phase.

In the case of a system constituted of three elements A, B and C, the liquid phase will be composed therefore of the simple species A, B and C, plus two associated species AC and BC (the AB species being disregarded caused to the weakness of the interaction [1]).

## 2. DESCRIPTION OF THE VAPOR PHASE

The activity of each constituent is given then for the mono-atomic vapor by the relation:

$$a_i^v = \frac{P_i}{P_i^0} \qquad (1)$$

And for the diatomic vapor by the relation:

$$a_i^v = \left(\frac{P_i}{P_i^0}\right)^{1/2} \qquad (2)$$

## 3. DESCRIPTION OF THE SOLID PHASE

The solid phases are the pure elements A or C, the defined compounds (AC and BC) and the pseudo-binary compound $(AC)_x(BC)_{1-x}$ or more easiness $A_xB_{1-x}C$ where $x$ is the composition in species AC. The free enthalpy of mixture from the species and by mole of species, be composed of: an ideal enthalpy by the formula:

$$\Delta_s^{id} G_s^S = RT\,[x \log x + (1-x) \log (1-x)] \qquad (3)$$

And an excess free enthalpy expressed in the model of pseudo-regular associated solution (PRAS) under the formula:

$$\Delta_s^{xc} G_s^S = \boldsymbol{a}\,x(1-x) \qquad (4)$$

With $\boldsymbol{a} = \boldsymbol{a}_H - T\boldsymbol{a}_S$ [2,3,4], represent the AC-BC interaction in the solid.

Designating by $\Delta^0 G_{AC}$ and $\Delta^0 G_{BC}$ the free enthalpies of formation of the two species respectively, the free enthalpy of the solid reported to the pure liquid elements is:

$$\Delta^m G_e^S = RT [x \operatorname{Log} x + (1-x) \operatorname{Log} (1-x)]$$
$$+ ax(1-x) + x\Delta^0 G_{AC} + (1-x)\Delta^0 G_{BC} \qquad (5)$$

Using the Gibbs-Duhem relation, one finds the expressions of the chemicals potential of the species in the solid, in the case of AC species one will have:

$$m_{AC}^S = RT \operatorname{Log} a_{AC}^S$$
$$= a_S(1-x)^2 + RT \operatorname{Log} x + \Delta^0 G_{AC} \qquad (6)$$

## 4. DESCRIPTION OF THE LIQUID PHASE

Writing $x_i$ ($i=1,2,3$) the molar concentrations of the elements and $y_i$ ($i=1,2,3,4,5$) the molar concentrations of the species, and adopting the model of sub-regular solution for the associated liquid, the free enthalpy of mixture from the species and by mole of species $\Delta_s^m G_s^L$, writes itself then:

$$\Delta_s^m G_s^L = \Delta_s^{id} G_s^L + \Delta_s^{xc} G_s^L \qquad (7)$$

Where $\Delta_s^{id} G_s^L$ is the ideal free enthalpy:

$$\Delta_s^{id} G_s^L = RT \sum_{i=1}^{5} y_i \operatorname{Log} y_i \qquad (8)$$

And $\Delta_s^{xc} G_s^L$ is the excess free enthalpy that writes itself to the second order as a polynomial of Redlich-Kistler under the formula:

$$\Delta_s^{xc} G_s^L = \left\{ \sum_{i=1}^{4} \sum_{j=i+1}^{5} y_i y_j [a_{i,j} + b_{i,j}(y_i - y_j)] \right\} \qquad (9)$$
$$- w y_1 y_2 y_3$$

$a_{i,j}$ and $b_{i,j}$ represents the binary interactions, and $w$ the ternary interaction A-B-C. These quantities depends linearly on the temperature according to the relations $a_{i,j} = a_{H,i,j} - T a_{S,i,j}$, $b_{i,j} = b_{H,i,j} - T b_{S,i,j}$, and $w = w_H - T w_S$.

The obtain nation of the liquid of species from the pure elements requires beforehand the formation of the associated species AC and BC, then the mixture of the set of species. While designating by $\Delta^d G_{AC}$ and $\Delta^d G_{BC}$ the free enthalpies of dissociation of species AC and BC, one will have:

$$\Delta^m G_s^L = \Delta_s^m G_s^L - y_4 \Delta^d G_{AC} - y_5 \Delta^d G_{BC} \qquad (10)$$

On the other hand, the free enthalpy of mixture of the liquid by mole of species also $\mu_{si}$ under the formula:

$$\Delta^m G_s^L = \sum_{i=1}^{5} y_i \mu_{si} \qquad (11)$$

While using the relation of Gibbs-Duhem $\left(\sum_{i=1}^{5} y_i d\mu_{si} = 0\right)$ one gets for the species $(k)$ :

$$\mu_{sk} = \Delta^m G_s^L + \frac{\partial(\Delta^m G_s^L)}{\partial y_k} \qquad (12)$$

Then clarifying the quantity $\Delta^m G_s^L$ according to the relation (10), and expressing the chemical potential according to the activity according to the relation $\mu_{sk} = RT \, \text{Log} \, a_{sk}$, and the activity according to the coefficient of activity according to $a_{sk} = y_k \gamma_{sk}$ :

$$RT \, \text{Log} \, \gamma_{sk} = \left[\sum_{i=1}^{4} \sum_{j=i+1}^{5} l_{i,j}(k) h_{i,j}(k)\right] - d_{4,k} \Delta^d G_{AC} - d_{5,k} \Delta^d G_{BC} \qquad (13)$$

Where:

$$l_{i,j}(k) = a_{i,j}[d_{k,i} y_j + d_{k,j} y_i - y_i y_j]$$
$$+ b_{i,j}[d_{k,i}(2 y_i y_j - y_j^2) - d_{k,j}(2 y_i y_j - y_i^2)$$
$$+ 2 y_i y_j (y_j - y_i)]$$

$$h_{i,j}(k) = -w[\ddot{a}_{k,1} y_2 y_3 + \ddot{a}_{k,2} y_1 y_3 + \ddot{a}_{k,3} y_1 y_2$$
$$- 2 y_1 y_2 y_3]$$

## 5. CONCENTRATIONS OF SPECIES IN LIQUID

$$x_i = \frac{1}{1 + y_4 + y_5}[y_i + (d_{1,i} + d_{3,i}) y_4 + (d_{2,i} + d_{3,i}) y_5] \qquad (14)$$

The report of the free enthalpies of mixture of the liquid by mole of elements $\Delta^m G_e^L$ and by mole of species $\Delta^m G_s^L$, equal to the report of the numbers of mole species $Ns$ and mole elements $Ne$, will be worth then:

$$\frac{\Delta^m G_e^L}{\Delta^m G_s^L} = \frac{Ns}{Ne} = \frac{1}{1 + y_4 + y_5} \qquad (15)$$

While expressing the quantity $\Delta^m G_e^L$ in function of the activities $a_{ei}$ of the elements, and $\Delta^m G_s^L$ in function of the activities $a_{si}$ of the species, according to the expressions:

$$\Delta^m G_e^L = RT \sum_{i=1}^{3} x_i \, \text{Log} \, a_{ei} \tag{16}$$

$$\Delta^m G_s^L = RT \sum_{i=1}^{5} y_i \, \text{Log} \, a_{si} \tag{17}$$

And using the relations (11), (12), and (14), one gets:
$$\begin{aligned} a_{s4} &= a_{s1}.a_{s3} \\ a_{s5} &= a_{s2}.a_{s3} \end{aligned} \tag{18}$$

Considering the expression $a_{si} = y_i \, s_i$ joining the chemical potential of every species to its activity, one gets the relations:

$$\begin{aligned} K_1 &= \frac{y_1 y_3}{y_4} = \frac{g_{s4}}{g_{s1} g_{s3}} \\ K_2 &= \frac{y_2 y_3}{y_5} = \frac{g_{s5}}{g_{s2} g_{s3}} \end{aligned} \tag{19}$$

Combining the relations (14), and introducing the coefficients $K_1$ and $K_2$, one expresses the concentrations of the associated species AC and BC in function of $y_3$ (the concentration of the species C):

$$y_4 = \frac{x_1 y_3 (1 - y_3)}{(1 - x_3)(y_3 + K_1)} \tag{20}$$

$$y_5 = \frac{x_2 y_3 (1 - y_3)}{(1 - x_3)(y_3 + K_2)} \tag{21}$$

And reporting in the relations (14), one gets the equation of the third degree in $y_3$:

$$a \, y_3^3 + b \, y_3^2 + c \, y_3 + d = 0 \tag{22}$$

Where: $a = 1 \; x_1 \; x_2$

$b = x_1 + x_2 \; x_3 + (K_1 + K_2) \; K_1 \, x_2 \; K_2 \, x_1$

$c = K_1 \, x_2 + K_2 \, x_1 \; (K_1 + K_2) \, x_3 + K_1 \, K_2$

$d = K_1 \, K_2 \, x_3$

The resolution of this equation allows us to get the concentration $y_3$ and with the relations (20) and (21) the compositions $y_4$ and $y_5$, and finally with the help of the relations (19) the concentrations $y_1$ and $y_2$.

# 6. QUANTITIES OF DISSOCIATION

$$\Delta^m H_e^L = (\Delta^0 H_{AC} + {}^f T^{AC} \Delta^0 S_{AC})/2 \tag{23}$$

$$\Delta^m S_e^L = (\Delta^0 S_{AC} + \Delta^f S_{AC})/2 \tag{24}$$

Where $\Delta^0 H_{AC}$, $\Delta^0 S_{AC}$ and $\Delta^f S_{AC}$ are the experimental thermodynamic quantities of formation and fusion of the compound CdTe and ${}^f T_{AC}$ its temperature of fusion.

One expresses therefore the free enthalpy of mixture of the liquid in $y_1 = y_3 = (1 - y_4)/2$ and at $T = {}^f T_{AC}$ according to the relation (13) and one deduces the quantities of dissociation:

$$\Delta^d H_{AC} = [y_3^2 \boldsymbol{a}_{H.1,3} + 2 y_3 y_4 \boldsymbol{a}_{H.ass} + 2 y_3 y_4 (y_3 - y_4) \boldsymbol{b}_{H.ass} - (1 + y_4)\Delta^m H_e^L ]/y_4 \tag{25}$$

$$\Delta^d S_{AC} = [y_3^2 \boldsymbol{a}_{S.1,3} + 2 y_3 y_4 \boldsymbol{a}_{S.ass} + 2 y_3 y_4 (y_3 - y_4) \boldsymbol{b}_{S.ass} - (1 + y_4)\Delta^m H_e^L - 2R \mathrm{Log} y_3 - R \mathrm{Log} y_4 ]/y_4 \tag{26}$$

Where: $\boldsymbol{a}_{H.ass} = (\boldsymbol{a}_{H.1,4} + \boldsymbol{a}_{H.1,4})/2$

$\boldsymbol{a}_{S,ass} = (\boldsymbol{a}_{S.1,4} + \boldsymbol{a}_{S.1,4})/2$

# 7. PARAMETERS OF INTERACTIONS

$$S_1 = \sum_1^N \left[ \sum_{i=1}^3 \left( \mathrm{Log} \frac{a_i^L}{a_i^V} \right)^2 \right] \tag{27}$$

$$S_2 = \sum_1^N \left( \frac{m_4^S - m_4^L}{m_4^S} \right)^2 \tag{28}$$

$$S_3 = \sum_1^N \left( \frac{m_3^S - m_3^L}{m_3^S} \right)^2 \tag{29}$$

$$S = S_1 + S_2 + S_3$$

Where $N$ is the number of available experimental points.

The calculation of the interaction parameters by minimization of the S function according to the simplex method of Nelder-Mead [5].

## 8. RESULTS AND DISCUTIONS

The results of the parameters for the binaries are reported in the table 1a. for the system Cd-Te and the table 1b. for Hg-Te.

These results show a very pronounced character of association for the binary Cd-Te where $y_{CdTe}(^fT_{CdTe}) = 0,9$ and of least importance for the binary Hg-Te where $y_{HgTe}(^fT_{HgTe}) = 0,4$. The parameters of interaction between the simple species Cd and Te of the first order are negligible compared to the parameters of the second order, what expresses a parabolic behavior of the free enthalpy in function of the concentrations in Cd and Te. On the other hand the parameters of interaction between a simple species and an associated species is therefore more important to the first order, what denotes a linear relation of the free enthalpy with the concentrations, then a pseudo-regular behavior of the liquid. As for the thermodynamic quantities of dissociation.

$y_{CdTe}(^fT_{CdTe}) = 0,9$

| (cal.mole$^{-1}$) | (cal.K$^{-1}$.mole$^{-1}$) | (cal.mole$^{-1}$) | (cal.K$^{-1}$.mole$^{-1}$) |
|---|---|---|---|
| $H._{Cd,Te} = 0,0$ | $S._{Cd,Te} = 6,203$ | $H._{Cd,Te} = -4506,0$ | $S._{Cd,Te} = -2,458$ |
| $H._{Cd,CdTe} = 27293,0$ | $S._{Cd,CdTe} = 19,97$ | $H._{Cd,CdTe} = 0,3$ | $S._{Cd,CdTe} = 0,2$ |
| $H._{Te,CdTe} = 5407,0$ | $S._{Te,CdTe} = 3,027$ | $H._{Te,CdTe} = 0,1$ | $S._{Te,CdTe} = 0,0$ |
| $^dH_{CdTe} = 20843,09$ | $^dS_{CdTe} = 3,884$ | | |

**Table 1.a**. Parameters and thermodynamic quantities for the binary Cd -Te.

$y_{HgTe}(^fT_{HgTe}) = 0,4$

| (cal.mole$^{-1}$) | (cal.K$^{-1}$.mole$^{-1}$) | (cal.mole$^{-1}$) | (cal.K$^{-1}$.mole$^{-1}$) |
|---|---|---|---|
| $H._{Hg,Te} = 1274$ | $S._{Hg,Te} = 4,111$ | $H._{Hg,Te} = 107$ | $S._{Hg,Te} = 0,991$ |
| $H._{Hg,HgTe} = 2251$ | $S._{Hg,HgTe} = -1,378$ | $H._{Hg,HgTe} = 0,1$ | $S._{Hg,HgTe} = 0,1$ |
| $H._{Te,HgTe} = 2167$ | $S._{Te,HgTe} = 0,91$ | $H._{Te,HgTe} = 0,1$ | $S._{Te,HgTe} = -0,1$ |
| $^dH_{HgTe} = 8038,34$ | $^dS_{HgTe} = 2,517$ | | |

**Table 1.b.** Parameters and thermodynamic quantities for the binary Hg -Te.

Reinforced by results, we supposed the interactions between two foreign species are pseudo-regulars.

| $H$.Cd,HgTe $= 0,0$ | $H$.Hg,CdTe $= -0,1$ | $H$.CdTe,HgTe $= 0,0$ | $H = 2,0$ | $H = 556,0$ |
|---|---|---|---|---|
| $S$.Cd,HgTe $= 0,2$ | $S$.Hg,CdTe $= 0,1$ | $S$.CdTe,HgTe $= 0,3$ | $S = 1,0$ | $S = 0.911$ |

**Table 2.** Parameters and thermodynamic quantities for the ternary Cd-Hg-Te.

These results show that the parameters of interaction in the liquid are very weak compared with the parameters in the binaries, what reinforces our hypothesis on the negligence of the interactions between foreign species. As for the parameter of interaction in the solid solution, the entropic term $S$ is negligible compared to the enthalpic term $H$ what proves an independent behavior of the temperature in this pseudo-binary solution.

## 9. PASES DIAGRAMS

To determine the phases diagrams, one looks for the minimum of the total free enthalpy of the two phases system:
$$\Delta G = q\Delta^m G_e^S + (1-q)\Delta^m G_e^L \tag{30}$$
Where $q$ represents the crystalline fraction in the system.

The minimum of $\Delta G$ is calculated by the method of Nelder-Mead, from where the results are in fig. 1, 2 and 3.

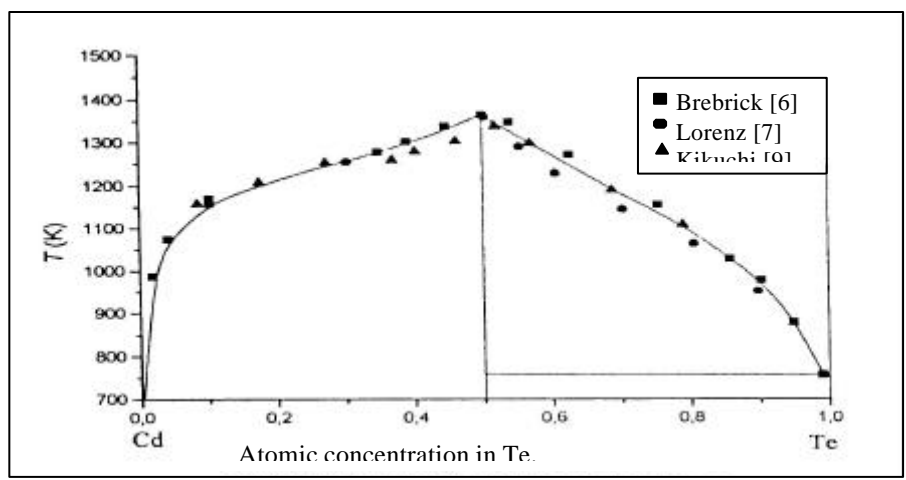

**Fig. 1.** Phases diagram for the system Cd-Te.

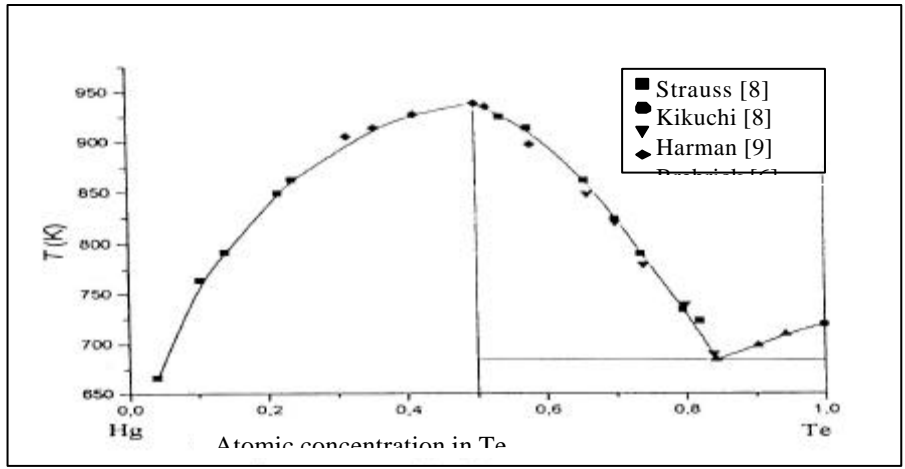

**Fig. 2.** Phases diagram for the system Hg-Te.

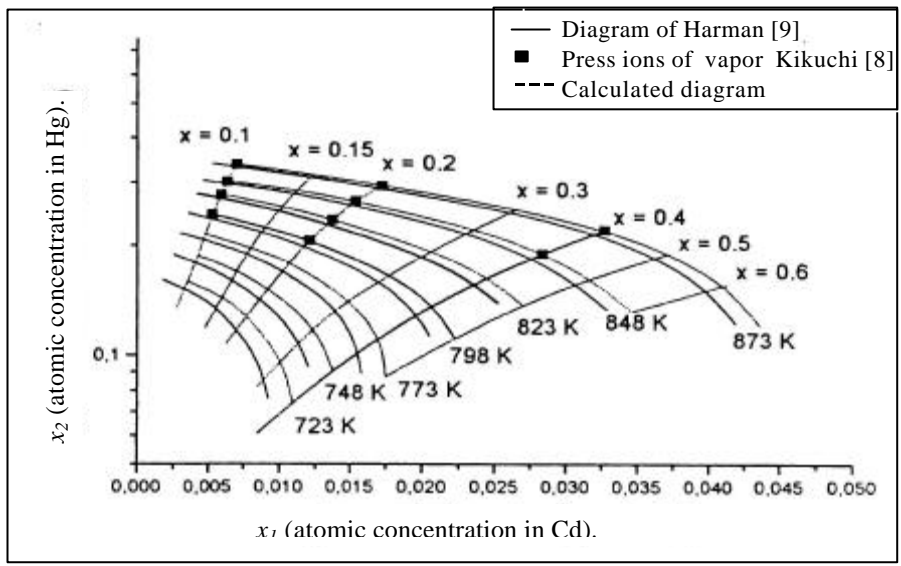

**Fig. 3.** Phases diagram for the system Cd-Hg-Te.

The calculated binary diagrams present a correct form that seems to interpolate the different experimental points in a very appropriate way. The point of eutexie for the system Hg-Te is foreseen at the temperature $T$

$x_{Te} = 0,84$.

As for the ternary diagram it presents a very similar form to the experimental diagram of Harman [9] used like reference, but nevertheless slightly removed, especially at low temperatures, and corner riche in Tellurium caused to the very limited number of available experimental points of vapor pressures.

## 10. CONCLUSION

The model of Sub-regular Associated Solution used to describe the systems Cd-Te, Hg-Te and Cd-Hg-Te , succeeds to a very appropriate phases diagrams, what permits an correct use of the thermodynamic quantities and the calculated parameters of interaction.